\documentclass[epj,nopacs]{svjour}
\usepackage{graphics}
\begin{document}
\title{Anisotropic flow fluctuations in hydro-inspired freeze-out model 
for relativistic heavy ion collisions}
\author{L.V.~Bravina\inst{2}, E.S.~Fotina\inst{1}, V.L.~Korotkikh\inst{1}, 
I.P.~Lokhtin\inst{1}, L.V.~Malinina\inst{1}, E.N.~Nazarova\inst{1}, 
S.V.~Petrushanko\inst{1}, A.M.~Snigirev\inst{1}, E.E.~Zabrodin\inst{1,2}}  
% etc
\institute {Skobeltsyn Institute of Nuclear Physics, Lomonosov Moscow State 
University, Moscow, Russia \and 
The Department of Physics, University of Oslo, Norway}
\abstract{ 
The LHC data on 
event-by-event harmonic flow coefficients measured in PbPb collisions at 
center-of-mass energy 2.76 TeV per nucleon pair are analyzed and 
interpreted within the HYDJET++ model. To compare the model results with 
the experimental data the unfolding procedure is employed. 
The essentially dynamical origin of the flow fluctuations in 
hydro-inspired freeze-out approach has been established. It is shown 
that the simple
modification of the model via introducing the distribution over spatial 
anisotropy parameters permits HYDJET++ to reproduce both elliptic and 
triangular flow fluctuations and related to it eccentricity fluctuations 
of the initial state at the LHC energy. 
 % end of PACS codes
}
%end of abstract
%
\titlerunning{Anisotropic flow fluctuations in hydro-inspired freeze-out 
model...}
\authorrunning{L.V. Bravina et al.}
\maketitle

\section{Introduction}
\label{sec_intro}

Azimuthal anisotropy of multi-particle production in relativistic heavy 
ion collisions is a powerful probe of collective properties of 
sub-nuclear matter created at extremely high densities and temperatures
(see, e.g., recent reviews~\cite{Heinz:2013th,Ritter:2014uca} and 
references therein). It is commonly described by the Fourier 
decomposition of the invariant cross section in a form: 
\begin{eqnarray}
\displaystyle
\label{eq:1}
E\frac{d^3N}{dp^3} &=&\frac{d^2N}{2\pi p_{\rm T}dp_{\rm T}d\eta} 
\nonumber \\
 &\times & \left\{
1+2\sum\limits_{n = 1}^\infty v_{\rm n}(p_{\rm T},\eta)  
\cos{ \left[ n(\varphi -\Psi_{\rm n}) \right] }
\right\} ~,
\end{eqnarray}
where $p_{\rm T}$ is the transverse momentum, $\eta$ is the 
pseudorapidity, $\varphi$ is the azimuthal angle with respect to the 
reaction plane $\Psi_{\rm n}$, and $v_{\rm n}$ are the Fourier 
coefficients. The observation of strong elliptic flow, which is the 
second harmonic, $v_2$, in heavy ion collisions at RHIC was argued as 
one of the main evidence for strongly-interacting partonic matter 
(``quark-gluon fluid'') formation~\cite{brahms,phobos,star,phenix}. 
At the LHC, a number of interesting measurements involving momentum and 
centrality dependencies of second and higher-order harmonic coefficients 
in PbPb collisions at $\sqrt{s}_{\rm NN}=2.76$~TeV have been done by 
ALICE~\cite{Aamodt:2010pa,ALICE:2011ab,Abelev2012:di}, 
ATLAS~\cite{ATLAS:2011yk,Aad:2012bu,Aad:2013xma,Aad:2014eoa,Aad:2014vba,Aad:2015lwa} and 
CMS~\cite{Chatrchyan:2012xq,Chatrchyan:2012ta,Chatrchyan:2012vqa,Chatrchyan:2013kba,CMS:2013bza} Collaborations. 
In particular, the event-by-event (EbyE) distributions of second, third and 
fourth harmonics of the anisotropic flow have been obtained~\cite{Aad:2013xma}.
Other important observations are the azimuthal anisotropy of 
jet~\cite{Aad:2013sla} and charmed meson~\cite{ALICE:2013xna,Abelev:2014ipa} 
yields in PbPb collisions, and elliptic $v_2$ and triangular $v_3$ flow of 
inclusive~\cite{Abelev:2012ola,Chatrchyan:2013nka,Aad:2014lta} 
and identified~\cite{Khachatryan:2014jra} hadrons in pPb collisions. 

In our previous study~\cite{Bravina:2013xla} the second and higher-order 
harmonics of inclusive and identified charged hadrons in PbPb collisions 
at $\sqrt{s}_{\rm NN}=2.76$ TeV were analyzed in the framework of 
HYDJET++ model~\cite{Lokhtin:2008xi}. It has been shown that the 
cross-talk of elliptic $v_2$ and triangular $v_3$ flow in the model 
generates both even and odd harmonics of higher order. This mechanism is 
able to reproduce the $p_{\rm T}$ and centrality dependencies of 
quadrangular flow $v_4$, and also the basic trends for pentagonal $v_5$ 
and hexagonal $v_6$ flows. Moreover, it reproduces also specific angular 
dihadron correlations including the so-called 
``ridge-effect''~\cite{Eyyubova:2014dha}. However, here we 
restricted ourselves to the analysis of the event-averaged harmonics 
$v_{\rm n}(p_{\rm T})$. In recent years, the study of anisotropic flow 
fluctuations has attracted much interest because of their direct 
connection with the geometry of the initial state of a relativistic 
heavy ion collision~\cite{Petersen:2013vca,Qian:2013nba,Huo:2013qma,Luzum:2013yya,Schenke:2013aza,Floerchinger:2014fta,Yan:2014afa,Bhalerao:2014xra,Song:2013gia,Jia:2014pza,Niemi:2015qia,Rybczynski:2015wva}.
In the present paper, therefore, we analyze the event-by-event 
distributions of the flow coefficients in PbPb collisions at the LHC 
within the HYDJET++ model. 

The paper is organized as follows. The flow fluctuations intrinsic 
to the HYDJET++ are discussed in Sec.~\ref{sec_fluct} Here the 
probability densities of both longitudinal and transverse flow components, 
as well as the flow  modulus, obtained at different collision centralities 
are shown to be nicely fitted to Gaussian. The fluctuations can be 
enhanced by the EbyE Gaussian smearing of the spatial anisotropy 
parameters of the model. Section~\ref{sec_unfold} describes the unfolding 
procedure proposed by the ATLAS Collaboration to get rid of the non-flow
fluctuations. This procedure is utilized in Sec.~\ref{sec_results} in
HYDJET++ calculations to compare the model results with the experimental
data on the same footing. The agreement with the data on fluctuations of
both elliptic and triangular flow is quite good. Conclusions are drawn 
in Sec.~\ref{sec_concl}. 

\section{Inherent flow fluctuations and eccentricity fluctuations in 
HYDJET++ model}
\label{sec_fluct}

Event generator HYDJET++ (the successor of HYDJET \cite{Lokhtin:2005px}) 
is the Monte-Carlo model of relativistic heavy ion collisions, which 
incorporates two independent components: the soft hydro-type state with 
preset freeze-out conditions, and the hard state resulting from the 
in-medium multi-parton fragmentation and taking into account jet 
quenching effect. The details of this model can be found in the HYDJET++ 
manual~\cite{Lokhtin:2008xi}. Its input parameters have been tuned 
to reproduce the experimental LHC data on various physical observables 
measured in PbPb collisions~\cite{Bravina:2013xla,Lokhtin:2012re},
namely, centrality and pseudorapidity dependence of inclusive 
charged particle multiplicity, transverse momentum spectra and 
$\pi^\pm \pi^\pm$ correlation radii in central PbPb collisions, 
momentum and centrality dependencies of elliptic and higher-order 
harmonic coefficients. 

In order to simulate higher azimuthal anisotropy harmonics, the following 
simple modification~\cite{Bravina:2013xla,Bravina:2013ora} has been 
implemented in the model. In original HYDJET++ 
version~\cite{Lokhtin:2008xi} the direction and strength of the 
elliptic flow are governed by two parameters. The spatial anisotropy 
$\epsilon(b)$ represents the elliptic modulation of the final freeze-out 
hyper-surface at a given impact parameter $b$, whereas the momentum 
anisotro\-py $\delta(b)$ deals with the modulation of flow velocity 
profile. 
Both $\delta (b)$ and $\epsilon (b)$ can be treated independently 
for each centrality, or (basic option of the model) can be related to 
each other through the dependence of the elliptic flow coefficient 
$v_2(\epsilon,\delta)$ obtained in the hydrodynamical 
approach~\cite{Wiedemann98}:
\begin{equation}
\label{v2-eps-delta1}
v_2(\epsilon, \delta) \propto \frac{2(\delta-\epsilon)}{(1-\delta^2)
(1-\epsilon^{2})}~.
\end{equation}

Then, due to the proportionality of $v_2(b)$ to the initial ellipticity
$\epsilon_0 (b)=b/2R_A$, where $R_A$ is the nuclear radius, the relation
between $\delta(b)$ and $\epsilon (b)$ takes the 
form~\cite{Lokhtin:2008xi}:
\begin{equation}
\label{v2-eps-delta2}
\delta = \frac{\sqrt{1+4B(\epsilon+B)}-1}{2B}~,~~~B=C(1-\epsilon^2)
\epsilon~,~~\epsilon=k \epsilon_0~,
\end{equation}
where two parameters $C$ and $k$ are independent on centrality and
should be obtained from the fit to the data.

Compared to the former transverse radius of the fireball, which 
reproduces the elliptic deformation
\begin{eqnarray}
\displaystyle
\label{eq:2}
& & R_{ell}(b,\varphi) = R_{f0}\frac{\sqrt{1-\epsilon^{2}(b)}}{\sqrt{1+
\epsilon(b)\cos2\varphi}}~,
\end{eqnarray}
the altered radius of the freeze-out hyper-surface in azimuthal plane 
takes into account triangular deformation as well:
\begin{eqnarray}
\displaystyle
\label{eq:3}
& & R(b,\varphi) =R_{ell}(b,\varphi) [1+\epsilon_{3}(b)
\cos[3(\varphi-\Psi_{3})]]~.
\end{eqnarray}
Here $\varphi$ is the spatial azimuthal angle of the fluid element relatively 
to the direction of the impact parameter. $R_{f0}$ is the model parameter 
which determines the scale of the fireball transverse size at freeze-out, 
and the new parameter $\epsilon_{3}(b)$ is responsible for the triangular 
spatial anisotropy. The event plane of the triangular flow, $\Psi_{3}$, 
is randomly oriented with respect to the plane $\Psi_{2}$, which is fixed 
to zero in the model calculations. This means that the elliptic and 
triangular flows are generated independently, in accordance with the 
experimental observations. Higher flow harmonics are not explicitly 
generated in the model, therefore these harmonics are absent if both 
$v_2$ and $v_3$ is absent.

It should be noted that although the azimuthal an\-iso\-tro\-py parameters 
$\epsilon(b)$, $\delta(b)$ and $\epsilon_3(b)$ are fixed at given impact 
parameter $b$, they define $v_{\rm n}(b)$ only after the averaging
over many events due to the inherent model fluctuations. The main source 
of the flow fluctuations in HYDJET++ is fluctuations of particle momenta 
and multiplicity. Recall, that the momentum-coordinate correlations in 
HYDJET++ for soft component are governed by collective velocities of the
fluid elements, and so the fluctuations in particle coordinates are 
reflected in their momenta. The fluctuations became stronger as resonance 
decays and (mini-)jet production are taken into account. An event 
distribution over collision impact parameter for each centrality class 
also increases such fluctuations. 

The detailed study of the event-by-event 
flow fluctuations is the subject of our present investigation. The 
possible further modification of HYDJET++ to match the experimental data 
on the flow fluctuations would be smearing of all three parameters, 
$\epsilon$, $\delta$ and $\epsilon_3$, at a given $b$.

\begin{figure*}
\begin{center}
\resizebox{1.\textwidth}{!}{%
\includegraphics{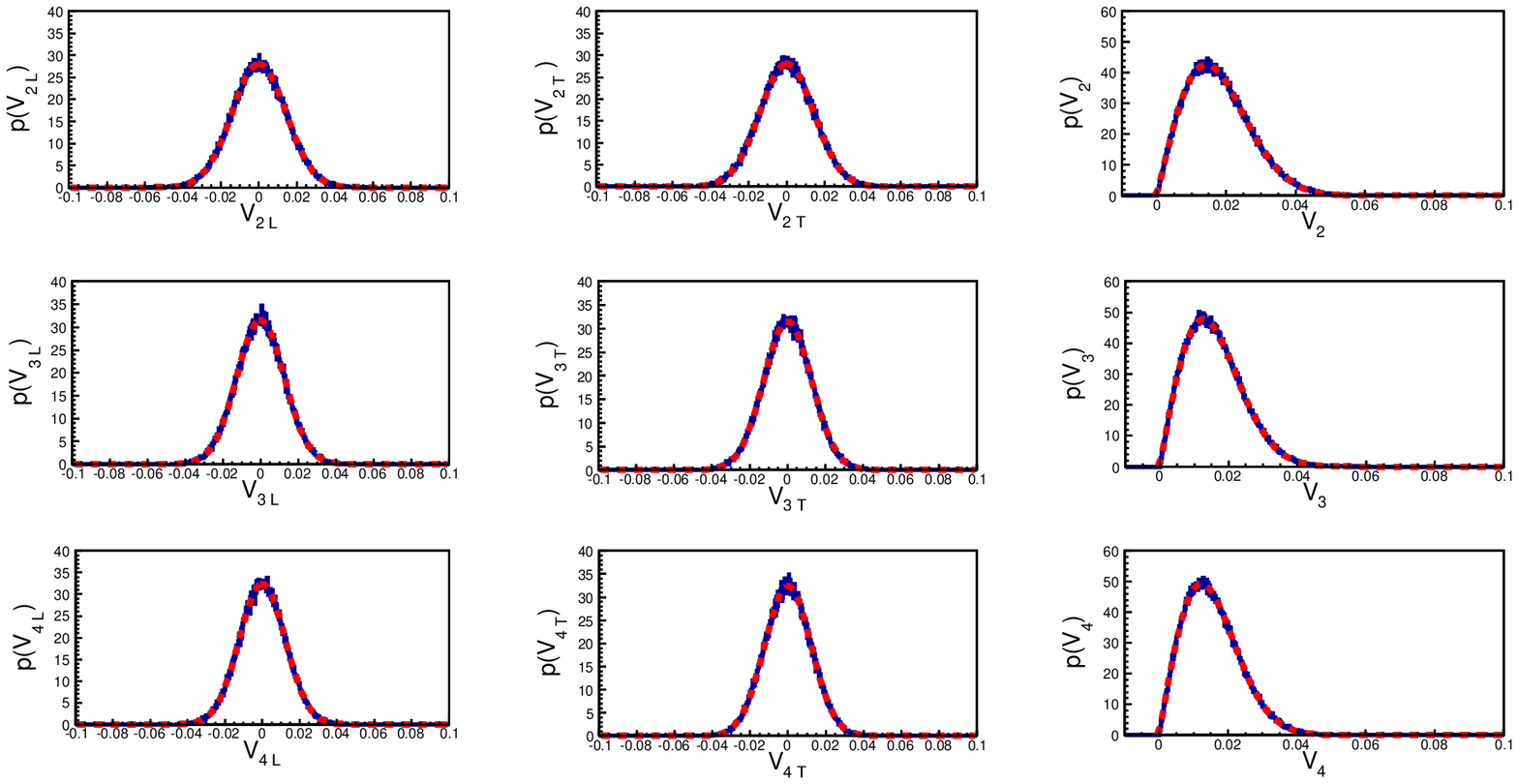}
}
\end{center}
 \caption{The probability density distributions for two components of 
the flow vector, $V_{\rm n T}$ and $V_{\rm n L}$, and for its modulus 
$V_n=|\vec{V}_n|,\ n = 2,3,4$ in the case of zero signal in HYDJET++ 
(central PbPb collisions at impact parameter $b=0$).
Dashed curves on the left and middle plots show two-dimensional fit of 
simulated HYDJET++ points to Eq.~(\ref{eq:4}); dashed curves on the 
right plots present the 1D-fit to Eq.~(\ref{eq:5}).} 
\label{fig1}
\end{figure*}

To get some notion for the inherent model fluctuations, we start with 
HYDJET++ simulations for the simplest case of central PbPb collisions 
$(b=0)$ in which all azimuthal anisotropy parameters $\epsilon(b)$, 
$\delta(b)$ and $\epsilon_3(b)$ are equal to zero. 
Figure~\ref{fig1} shows the probability densities both for each component 
of the flow vector $\vec{V}_n$ and for its modulus $V_n=|\vec{V}_n|$, 
$n=2,3,4$. 
\footnote{Via $\vec{V}_n$ we denote the flow vector determined on 
EbyE basis, while the standard definition $v_n = \langle
\cos{\left[ n (\varphi - \Psi_{\rm n}) \right]} \rangle$ is reserved for
the flow components obtained by the averaging over all particles in an
event and over all events in the data sample.}  
In this ``fluctuation-only'' scenario the probability densities 
of $\vec{V}_n$ are well described by two-dimensional (2D) Gaussian
functions~\cite{Voloshin:2007pc,Bhalerao:2006tp}:
\begin{eqnarray}
\displaystyle
\label{eq:4}
& & p(\vec{V}_n)=\frac{1}{2\pi \sigma^2_{n}} 
\exp[- \vec{V}_n^2/(2\sigma^2_{n})]~,
\end{eqnarray}
whereas the probability densities of $V_n$ have the forms of
one-dimensional (1D) Gaussians
\begin{eqnarray}
\displaystyle
\label{eq:5}
& & p(V_n)=\frac{V_n}{ \sigma^2_{n}} \exp[- V_n^2/(2\sigma^2_{n})]~,
\end{eqnarray}
which are obtained from Eq.(\ref{eq:4}) by integration over the azimuthal 
angle.

These distributions are characterized by a single parameter $\sigma_{n}$ 
only, which regulates both the modulus mean $\langle V_n \rangle$ and 
the width 
$\sigma_{V_n}=\sqrt{\langle V_n^2 \rangle -\langle V _n \rangle ^2}$ as
\begin{eqnarray}
\displaystyle
\label{eq:6}
& & \langle V_n \rangle =\sqrt{\frac{\pi}{2}} ~\sigma_{n}~,\\ 
\label{eq:7}
& & \sigma_{V_n}=\sqrt{2-\frac{\pi}{2}}~ \sigma_{n}~,\\
\label{eq:8}
& & \frac{\sigma_{V_n}}{\langle V_n \rangle }=\sqrt{\frac{4}{\pi} 
- 1}~=~0.523~.
\end{eqnarray}

In HYDJET++ the value of this fitting  parame\-ter, Gaussian width 
$\sigma_{n}$, appears to be unique for all harmonics with a good 
enough accuracy: 
$\sigma_{2}\simeq \sigma_{3} \simeq \sigma_{4} \approx 0.013$. It depends 
on a number of model parameters, which were already fixed. The main 
regulator of $\sigma_{n}$ is the mean multiplicity, which determines the 
variation of $\sigma_{n}(b)$ with centrality. 

Recall, that in all Bjorken-like models with cylindrical 
parameterization the azimuthal anisotropy of the freeze-out surface
transforms into the azimuthal anisotropy of particle momentum 
distribution proportionally to a term $(p_T \sinh{Y_T}/T) \cos{(\phi -
\varphi)}$ \cite{Lokhtin:2008xi}, arising in scalar product of 4-vectors
of particle momentum and flow velocity of the fluid element. Here $\phi$
is the azimuthal angle of the fluid element, $\varphi$ is the particle
azimuth, $T$ is the freeze-out temperature and $Y_T$ is the transverse 
flow rapidity, respectively. The pre-factor before the cosine controls 
the azimuthal angle structure of particle spectrum and its inverse 
characterizes the fluctuation width squared. We have also verified 
numerically that at fixed mean multiplicity in selected $p_T$ window
the width of $\sigma_{n}(p_T)$ is approximately proportional to the 
factor $\sqrt{T/(p_T \sinh{Y_T^{\rm max})}}$, where $Y_T^{\rm max}$ is 
the maximal transverse flow rapidity. 

The ``true'' direction of the flow vector $\vec{V}_n$ in HYDJET++ for any 
azimuthal harmonic is pre-defined in each event. Therefore, 
$V_{\rm n L}$ can be calculated as a longitudinal component of the 
actual flow vector along this known direction, and $V_{\rm n T}$ --- 
as a transverse component of $\vec{V}_n$ perpendicular to the
longitudinal one. 
As shown in Fig.~\ref{fig1}, the mean values of the flow components 
are zero, $ \langle V_{\rm nL} \rangle = \langle V_{\rm nT} \rangle = 0,
\ n = 2,3,4$, and, therefore, $v_2 = v_3 = v_4$ as it should be in a 
trivial case of the flow absence in a perfectly central collision.

\begin{figure*}
\begin{center}
\resizebox{1.\textwidth}{!}{%
\includegraphics{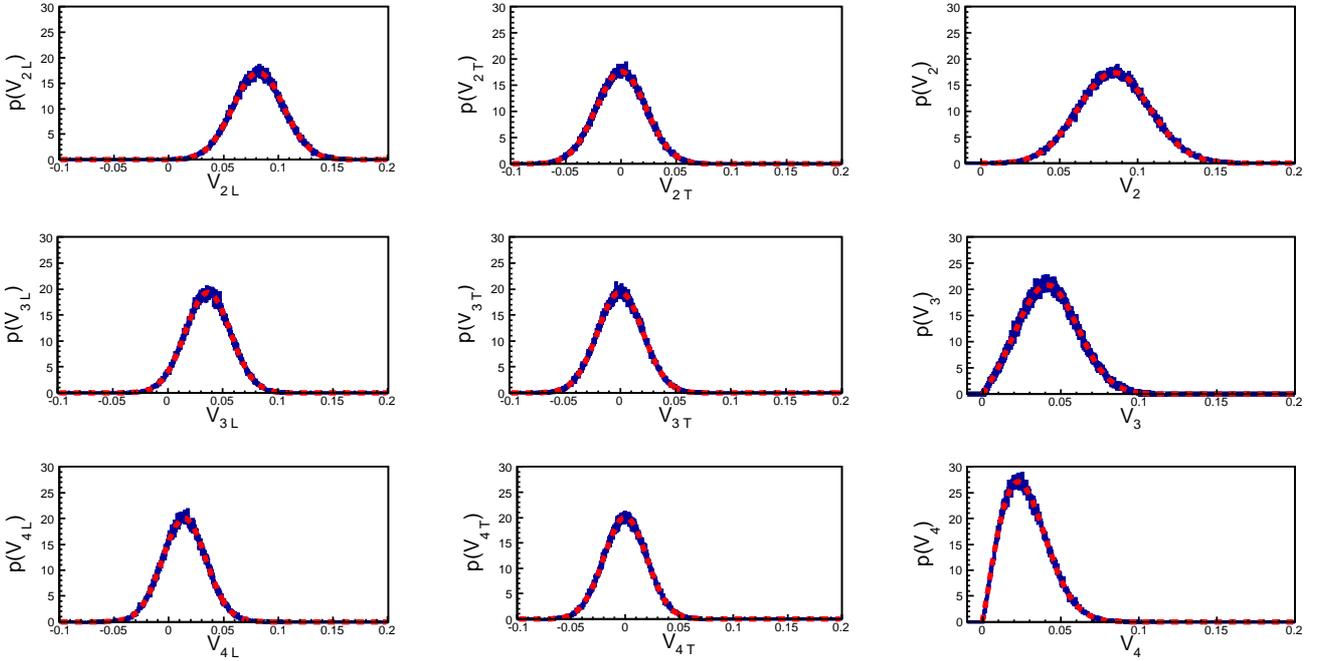}
}
\end{center}
\caption{The same as Fig.~\protect\ref{fig1} but for non-zero signal 
in HYDJET++ (20-25 \% centrality of PbPb collisions).
Dashed curves in the left, middle and right columns indicate the fit 
of simulated HYDJET++ points to Eqs.~(\ref{eq:9}), (\ref{eq:10}) and 
(\ref{eq:11}), respectively.} 
\label{fig2}
\end{figure*}

Figure~\ref{fig2} demonstrates the probability densities both for each 
component of the flow vector $\vec{V}_n$ and for its modulus 
$V_n=|\vec{V}_n|,\ n=2,3,4$ in the case of non-zero flow vector 
$\langle V_{\rm n L} \rangle $ in PbPb collisions at centralities $ 20 - 25\% $.
Instead of distributions (\ref{eq:4}) and (\ref{eq:5}) we get
here~\cite{Voloshin:2007pc,Bhalerao:2006tp}
\begin{eqnarray}
\displaystyle
\label{eq:9}
p(V_{\rm n L}) &=& \frac{1}{\sqrt{2\pi \sigma^2_{n}(b)}} \exp\left[
-\frac{(V_{\rm n L}- \langle V_{\rm n L }\rangle )^2}{2\sigma^2_{n}(b)}
\right]\ , \\
\label{eq:10}
p(V_{\rm n T}) &=& \frac{1}{\sqrt{2\pi \sigma^2_{n}(b)}} \exp\left[
-\frac {(V_{\rm n T})^2}{2\sigma^2_{n}(b)}\right] \ , \\
\label{eq:11}
\nonumber
p(V_n) &=& \frac{V_n}{ \sigma^2_{n}(b)} \exp\left[- \frac{V_n^2+ 
\langle V_{\rm n L}\rangle ^2}{2\sigma^2_{n}(b)}\right] 
I_0\left(\frac{V_n \langle V_{\rm n L}\rangle }{\sigma^2_{n}(b)}\right)
\ , \\
& &
\end{eqnarray}
where $I_0$ is the modified Bessel function of the first kind zero 
order. Both the width $\sigma_{V_n}=\sqrt{\langle V_n^2 \rangle -
\langle V_n \rangle ^2}$ and the modulus mean $\langle V_n \rangle$ are 
controlled by the true value of the flow vector $\langle V_{\rm n L} 
\rangle $ and the width $\sigma_{n}(b)$, but cannot be cast
analytically as functions of $\langle V_{\rm n L} \rangle $ and 
$\sigma_{n}(b)$. Note also that $\langle V_n \rangle$ is not equal to  
$\langle V_{\rm n L} \rangle $ exactly, and the azimuthal anisotropy 
parameters $\epsilon(b)$, $\delta(b)$ and $\epsilon_3(b)$ have been 
tuned earlier at a given impact parameter $b$ in such a way that the 
value of $\langle V_{\rm n L} \rangle $ extracted from the distribution 
(\ref{eq:9}) reproduces just the experimentally observed value of 
$v_{\rm n}(b)$ entering in Eq.(\ref{eq:1}).
Similar to the non-flow case, presented in Fig.~\ref{fig1}, the
widths $\sigma_n$ of the longitudinal and transverse distributions
shown in Fig.~\ref{fig2} are approximately the same, $\sigma_2 \simeq
\sigma_3 \simeq \sigma_4 \approx 0.02$, but the distributions become 
broader. Also, the maxima of $p(V_{\rm n L})$ and $p(V_n)$ distributions
are shifted towards zero with rising harmonic number $n$, indicating
that $v_2 > v_3 > v_4$ at this centrality. 
Surely, it is interesting to compare our inherent model probability 
densities, obtained without any additional special parameters for the
azimuthal fluctuations, with the experimental data.

We have also considered including of additional ``eccentricity'' 
fluctuations in HYDJET++ model. The simplest modification for this 
purpose is to introduce event-by-event Gaussian smearing of the spatial 
anisot\-ro\-py parameters $\epsilon(b)$ and $\epsilon_3(b)$ with the 
widths proportional to its ``unsmeared'' values. The coefficients of 
this proportionality are independent on event centrality and tuned 
to fit the data. 
Both model versions, with and without the smearing, are employed 
for the extraction of the flow fluctuations and comparison with the
available experimental data.

\section{Unfolding procedure for flow fluctuations analysis}
\label{sec_unfold}

Unfortunately, the direct comparison of the model distributions given
by Eqs.~(\ref{eq:9}-\ref{eq:11}) with the corresponding experimental 
data is impossible. The EbyE distributions of anisotropic flow harmonics 
have been obtained by the ATLAS Collaboration~\cite{Aad:2013xma} for the 
distribution of the modulus of the flow vector by application of the 
so-called ``unfolding procedure''. 
The goal of the unfolding procedure was to extract the 
``true'' flow vector from the observed one by excluding the influence of 
non-flow effects, such as resonance decays and jet fragmentation, as well 
as the finite event multiplicity effect. 
Therefore, in what follows we will show our results before and after the 
unfolding to be adequate. In order to employ the EbyE unfolding procedure 
for simulated events, the analysis method from~\cite{Aad:2013xma} was 
utilized.  

\begin{itemize} 

\item The EbyE distributions of charged particles in PbPb collisions at 
$\sqrt{s}_{\rm NN}=2.76$ TeV with $ p_{T} > 0.5$ GeV/$c$ and 
$ | \eta | < 2.5 $ are used as input distributions. Fourier decomposition 
of the azimuthal distribution is rewritten as

\begin{eqnarray}
\displaystyle
\label{eq:12}
\frac{dN}{d\varphi} &\propto&  1 + 2 \sum^{\infty}_{n=1}  
V^{\rm obs}_{\rm n} \cos{\left[ n(\varphi - \Psi^{\rm obs}_{\rm n})\right]} = 
\nonumber \\
&=&
1 + 2 \sum^{\infty}_{n=1} ( V^{\rm obs}_{\rm n,x} \cos{n\varphi} + 
V^{\rm obs}_{\rm n,y} \sin{n\varphi}) \ ,
\end{eqnarray}
where $ V^{\rm obs}_{\rm n}$ is the magnitude of the observed per-particle 
flow vector and $ \Psi^{\rm obs}_{\rm n}$ is the event plane angle. 
\item The single-particle EbyE distributions are constructed and used in 
our unfolding procedure:

\begin{eqnarray}
\displaystyle
\label{eq:13}
\nonumber
& & 
V^{\rm obs}_{\rm n} = \sqrt{(V^{\rm obs}_{\rm n,x})^{2} + 
(V^{\rm obs}_{\rm n,y})^{2}} \ , \\
& &
V^{\rm obs}_{\rm n,x} = V^{\rm obs}_{\rm n} \cos{n\Psi^{\rm obs}_{\rm n}} 
= \langle  \cos{n\varphi} \rangle \ , \\
\nonumber 
& & 
 V^{\rm obs}_{\rm n,y} = V^{\rm obs}_{\rm n} \sin {n\Psi^{\rm obs}_{\rm n}} 
= \langle  \sin{n\varphi} \rangle \ . 
\end{eqnarray}
The averaging in last two equations of (\ref{eq:13}) is performed over 
all hadrons in a single event.
As was shown in~\cite{Aad:2013xma}, the distribution obtained after the 
unfolding procedure did not depend on the method applied to obtain 
$V^{\rm obs}_{\rm n}$. Thus, the single-particle method can be used for 
our study. 

\item The response function is constructed using the ``two sub-event 
method'' (2SE), namely, the charged particles are divided into two 
sub-events with $\eta<0$ and $\eta>0$. The smearing effects are 
estimated by the difference of the flow vectors between the two 
sub-events, for which the flow signal cancels. This distribution is 
fitted to the Gaussian with the width $\delta_{\rm 2SE}$ determined 
mainly by the finite multiplicity effect and the non-flow contributions. 
It was shown in~\cite{Aad:2013xma} that the response function
constructed by such a procedure can be expressed as
\begin{eqnarray}
\displaystyle
\label{eq:14}
\nonumber
& & 
p ( V^{\rm obs}_{\rm n} \vert V_{\rm n} ) \propto V^{\rm obs}_{\rm n} 
\exp \left[ - \frac{( V^{\rm obs}_{\rm n})^{2} + V^{2}_{\rm n}}
{2\delta^{2}} \right] 
I_{0}\left( \frac{V^{\rm obs}_{\rm n}V_{\rm n}}{\delta^{2}} \right) 
\ . \\
& &   
\end{eqnarray}
Here $ \delta = \delta_{\rm 2SE}/2$ because we use the full-event 
$V^{\rm obs}_{\rm n}$ distribution as an input, and $\delta_{\rm 2SE}$ 
is the width obtained from the difference between the EbyE per-particle 
flow vectors of the two sub-events. 

\item The constructed response function is used to obtain the unfolding 
matrix:  
\begin{eqnarray}
\displaystyle
\label{eq:15}
& &     M_{ij}^{\rm iter} = \frac{A_{ji} c_{i}^{\rm iter}}{\Sigma_{m,k} 
A_{mi} A_{jk} c_{k}^{\rm iter}}, 
\nonumber \\
& & \widehat{c}^{\rm iter+1} =  \widehat{M}^{\rm iter} \widehat{e} ,  
A_{ji} = p(e_{j} \vert c_{i})~,
\end{eqnarray}
where $A_{ji}$ is the response function between $e_{j} = V^{\rm obs}_{\rm n}$ 
(``effect'') and $c_{i} = V_{\rm n}$ (``cause''). The true $V_{\rm n}$ 
distribution (cause ``c'') is obtained from the measured
$ V^{\rm obs}_{\rm n}$ distribution (effect ``$e$'') using an iterative 
algorithm. Then the Bayesian unfolding procedure is performed by means of
the RooUnfold package~\cite{Adye}.

\end{itemize}
 
The difference between the $\delta_{\rm 2SE}$ and $\sigma_n$
arises mostly due to the dynamical flow fluctuations. Therefore,
the EbyE unfolding analysis excludes the effects related to 
$ \delta_{\rm 2SE}$, and leaves the genuine flow fluctuations.
 
The application of Bayesian unfolding method for the anisotropic flow 
analysis was checked with heavy ion event generators in~\cite{JiaMoham}. 
It was shown that the restored density distributions were 
able to reproduce the input $V_{n}$ distributions. The non-flow 
effects were estimated by using the HIJING event generator~\cite{hijing} 
with and without the implementation of the flow signal, and were found 
to be of the order of statistical errors. 
In case of the AMPT model~\cite{ampt}, which includes both flow and 
non-flow fluctuations, the difference between the ``generated" and
``unfolded" flow harmonics in semiperipheral Au+Au collisions at RHIC 
was found to be small for elliptic flow and significantly increasing in 
the tails for triangular and quadrangular flows.
Fluctuations originated from the finite multiplicity effect can be 
evaluated under the assumption of Gaussian multiplicity 
distribution~\cite{JiaMoham}: 
 \begin{eqnarray}
\displaystyle
\label{eq:16}
& &  \delta_{n} = \sqrt{\Big\langle \frac{1}{2N} \Big\rangle} \simeq 
\sqrt{ \frac{1}{2 \langle N \rangle} \Big{[} 1+\Big{(} 
\frac{\sigma_{N}}{\langle N \rangle} \Big{)}^{2} \Big{]}}~, 
\end{eqnarray}
where $\langle N \rangle$ and $\sigma_{N}$ are the mean value and the
width of the distribution, respectively.

\section{Comparison of HYDJET++ simulations with LHC data}
\label{sec_results}

\begin{figure*}
\begin{center}
\resizebox{1.\textwidth}{!}{%
\includegraphics{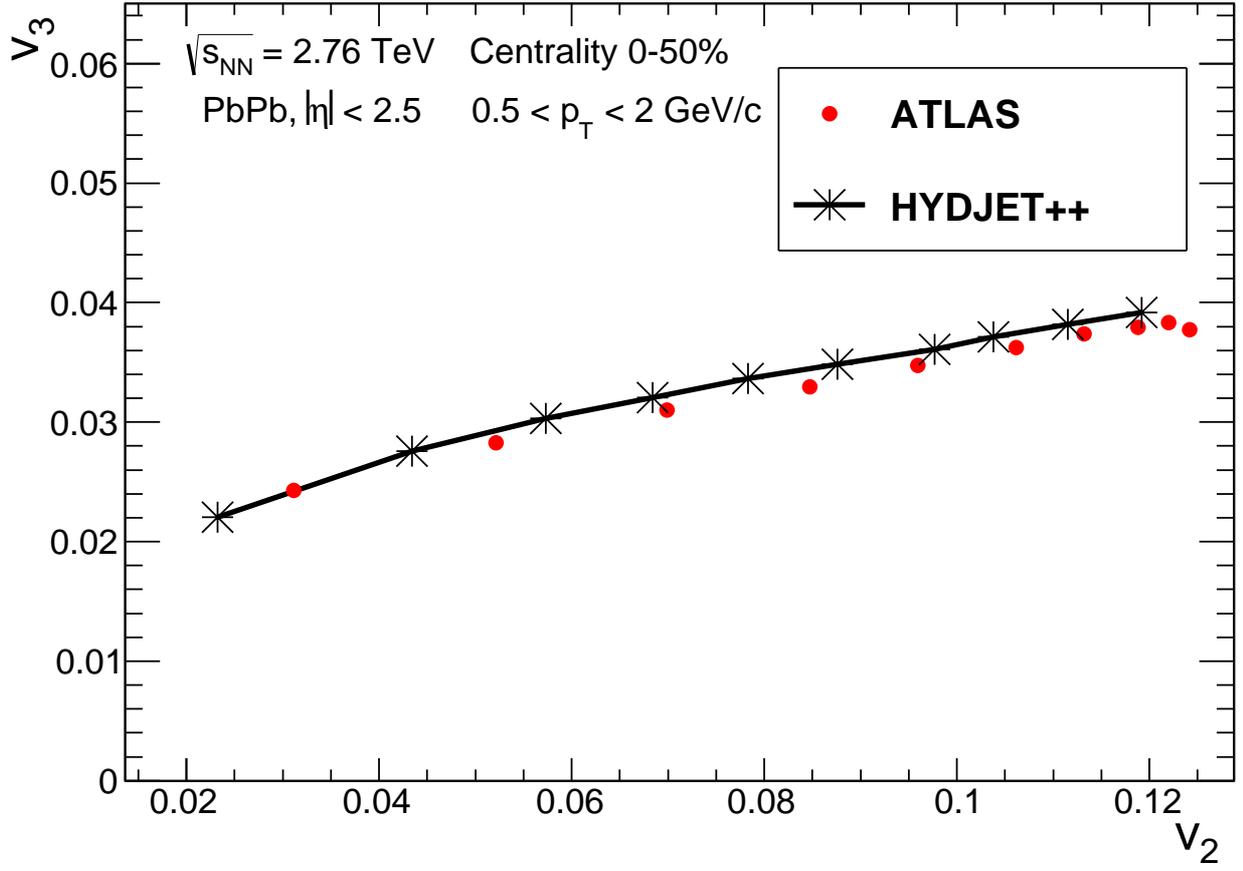}
}
\end{center}
\caption{The correlation between the event-averaged elliptic flow $v_2$ 
and triangular flow $v_3$ of charged hadrons at transverse momentum 
$0.5 < p_{\rm T} < 2$ GeV/$c$ and pseudo-rapidity $|\eta|<2.5$ 
for ten 5\%-centrality intervals in the centrality range $0-50 \%$ of PbPb 
collisions at $\sqrt s_{\rm NN}=2.76$ TeV. Closed circles denote ATLAS 
data from~\cite{Aad:2015lwa}, asterisks represent HYDJET++ events.
Line is drawn to guide the eye.} 
\label{fig3}
\end{figure*}

At first we have checked that HYDJET++ reproduces well the experimentally 
measured correlation between the event-averaged 
elliptic and triangular flow coefficients. The results of model
simulations are plotted onto the ATLAS data~\cite{Aad:2015lwa} in 
Fig.~\ref{fig3}. One can see that the calculations and the data agree 
well within the 7\% accuracy limit.

\begin{figure*}
\begin{center}
\resizebox{0.85\textwidth}{!}{%
\includegraphics{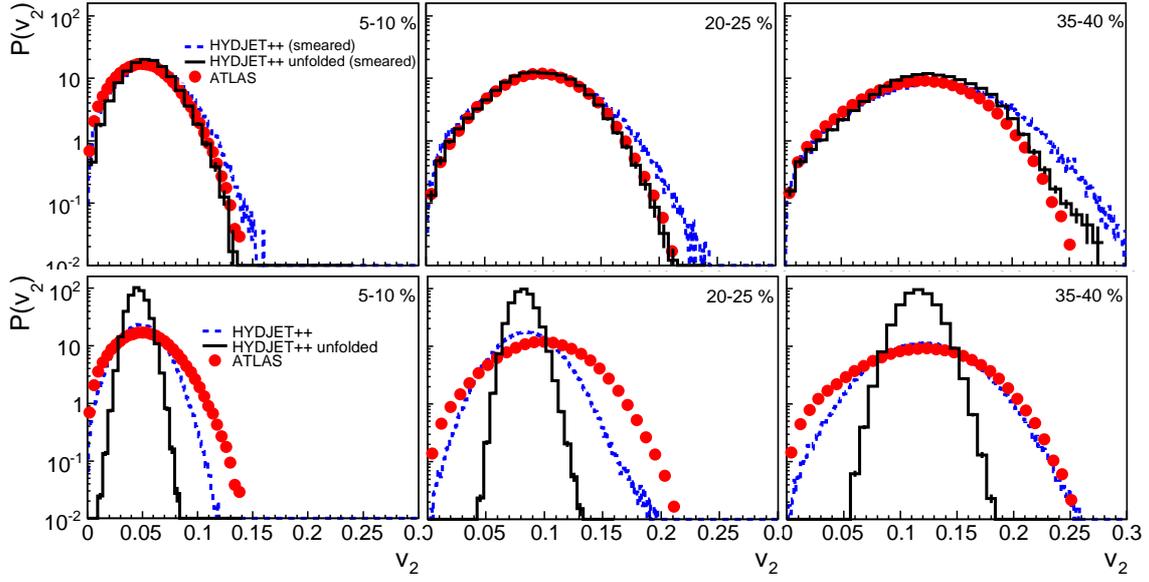}
}
\end{center}
\caption{The probability density distributions of elliptic flow $V_2$ in 
three centrality intervals $5-10 \%$ (left), $20-25 \%$ (middle) and 
$35-40 \%$ (right). Dashed and solid histograms present the results for 
simulated HYDJET++ events before and after the unfolding procedure, 
respectively. The top/bottom row shows the model results with/without 
the additional smearing of spatial anisotropy parameters. 
The closed points are ATLAS data from~\cite{Aad:2013xma}.} 
\label{pv2}
\end{figure*}

\begin{figure*}
\begin{center}
\resizebox{0.85\textwidth}{!}{%
\includegraphics{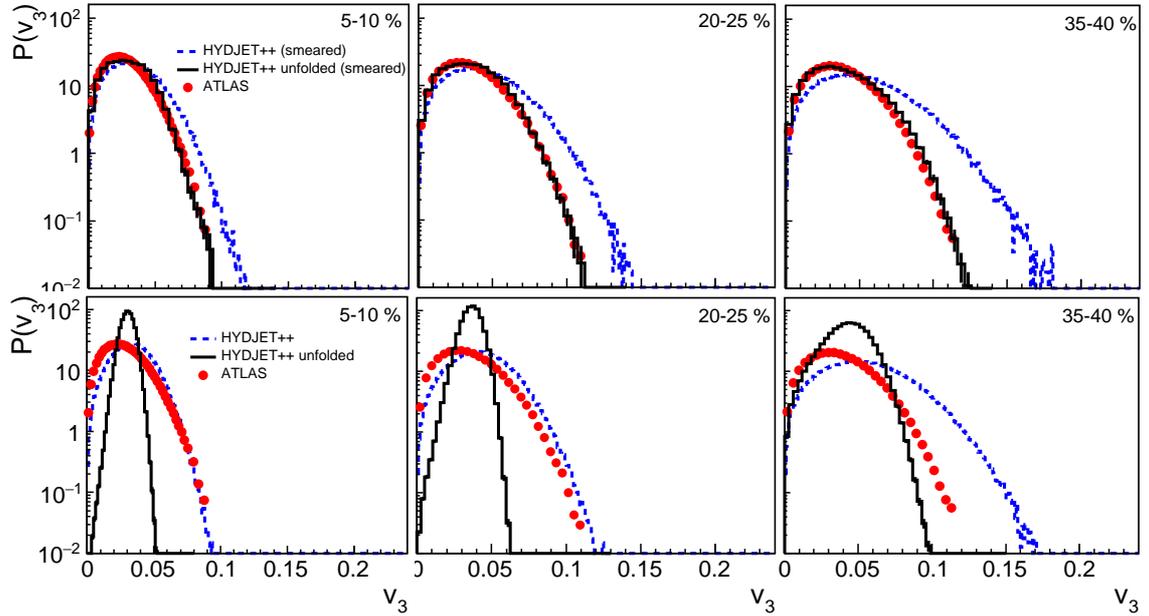}
}
\end{center}
\caption{The same as Fig.~\protect\ref{pv2} but for the triangular 
flow $V_3$ in three centrality intervals. } 
\label{pv3}
\end{figure*}

Then, we consider the anisotropic flow fluctuations. The simulations and 
analysis were performed for three  centrality intervals, namely, $5-10 \%$, 
$20-25 \%$ and $35-40 \%$, and for two settings of the HYDJET++ model: 
(i) without and (ii) with the additional smearing of spatial anisotropy 
parameters $\epsilon(b)$ and $\epsilon_3(b)$, see Sec.~\ref{sec_fluct}. 
The results for $p(V_2)$ and $p(V_3)$ are presented in Fig.~\ref{pv2} 
and Fig.~\ref{pv3}, respectively, and listed in Table~1. Both 
figures indicate  that the original version of HYDJET++ (without the 
smearing of $\epsilon(b)$ and $\epsilon_3(b)$) already includes some 
dynamical fluctuations due to radial flow. 
Moreover, the mean values of $\langle V_n \rangle$ and widths of 
the distributions $\sigma_{V_n}$ in the default version of HYDJET++ are
quite close to the measured ones in collisions with centralities up to 
25\% for the triangular flow and up to 45\% for the elliptic flow, see
Table~1. Although the agreement can be further improved by rescaling of
$\langle V_n \rangle$ to match the data, such approach would be 
completely misleading. Implementation of the unfolding procedure 
clearly demonstrates in Figs.~\ref{pv2},\ref{pv3} that the initial
distributions become more narrow. Thus, the intrinsic fluctuations
appear to be too weak to match the experimental data~\cite{Aad:2013xma}. 
On the other hand, simple 
modification of the model via introducing the normal distribution 
over the spatial anisotropy parameters allows us to reproduce the 
measured event-by-event fluctuations of elliptic and triangular flow, 
the distribution widths and the event-averaged values of 
$\langle V_2 \rangle$ and $\langle V_3 \rangle$. 
The distributions $p(V_n)$ obtained with the set of the smeared
out parameters are broader in the tails compared with the experiment.
Unfolding makes it narrower. Here the difference between the ``initial"
and ``unfolded" spectra are not so dramatic although still noticeable 
in contrast to that at RHIC energy, calculated in \cite{JiaMoham} 
within the AMPT model.
The two 
additional parameters of the model, appeared in this case, are the
coefficients of proportionality between the Gaussian widths of the 
distributions $p(\epsilon)$ and $p(\epsilon_3)$ and their ``unsmeared'' 
values. These two coefficients are fixed to fit the data on $p(V_2)$ 
and $p(V_3)$, respectively, for only one arbitrary centrality, whereas 
for other centralities they are the same. 

It is worth noting, that such a simple modification of the model also 
increases EbyE fluctuations for higher order harmonics $v_n$ ($n>3$), 
which arise in HYDJET++ due to the presence of elliptic $v_2$ and 
triangular $v_3$ flows, and its interference. However, significant 
sensitivity of high harmonic values on their extraction methods makes 
the direct comparison of our simulations with the data even more tricky 
than for $v_2$ and $v_3$. For example, the centrality dependence of 
quadrangular flow $v_4$ measured by event plane and two-particle cumulant 
methods is significantly weaker than that of $v_4$ measured by Lee-Yang 
zero method due to large non-flow contribution and increase of the flow 
fluctuations in more central events. Since HYDJET++ was tuned to fit the 
$p_{\rm T}$ -dependence of $v_4\{LYZ\}$, it underestimates $v_4$ extracted 
by the event plane or two-particle cumulant methods in (semi)central 
collisions~\cite{Bravina:2013xla}. We plan to study the event-by-event 
fluctuations of higher order flow harmonics in the future. 

\begin{figure*}
\begin{center}
\resizebox{0.85\textwidth}{!}{%
\includegraphics{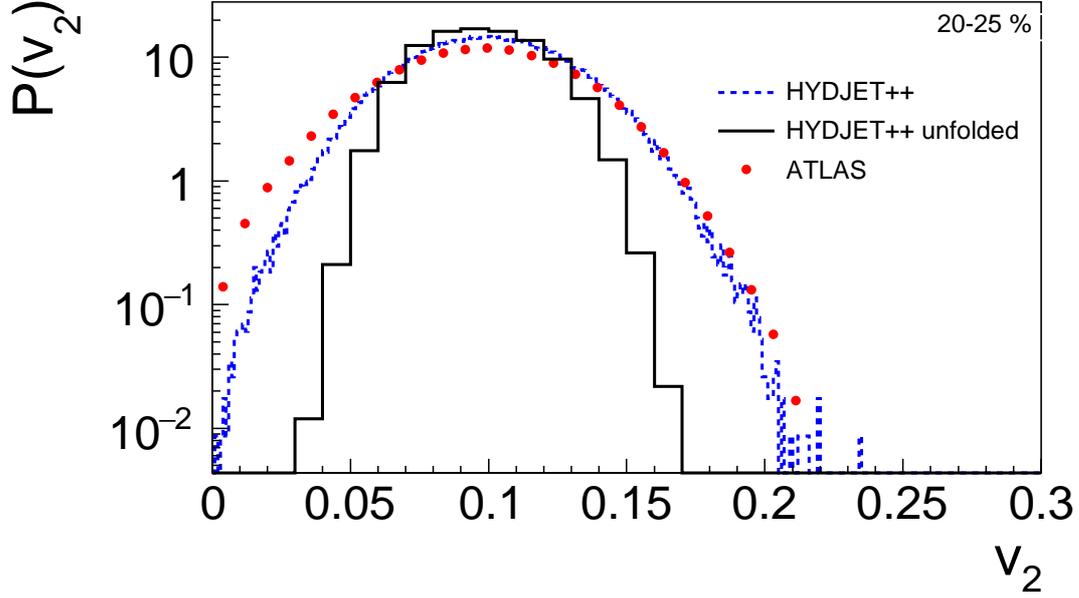}
}
\end{center}
\caption{The probability density distributions of elliptic flow $V_2$ in
centrality interval $20-25 \%$ obtained with uniformly smeared parameters
$\epsilon$ and $\delta$. Dashed and solid histograms present
the results for simulated HYDJET++ events before and after the unfolding 
procedure, respectively. ATLAS data are shown by full circles.
}
\label{uniform}
\end{figure*}

Few important issues should be clarified still. We utilized 
normal smearing of the parameters $\epsilon(b), \delta(b)$ and
$\epsilon_3(b)$ in the modified version of the event generator. What
will happen if the parameters are smeared out with respect to a
non-Gaussian distribution? To check this possibility we opted for a
uniform distribution of the key parameters within the interval
$\pm \sigma_{\epsilon}$ to ensure the same mean and width 
values. The distribution
$p(V_2)$ is displayed in Fig.~\ref{uniform} for centrality bin $20-25 \%$,
where the signal heavily dominates over the fluctuations. Interestingly
enough, the generated distribution is very close to the ATLAS unfolded
curve everywhere, but in the low-$V_2$ range. Unfolded HYDJET++ spectrum,
however, is narrow. It resembles the implemented rectangular-shaped 
$V_2$-dis\-tri\-bu\-tion with some rounding of the shoulders because of 
the intrinsic model fluctuations. 

\begin{figure*}
\begin{center}
\resizebox{0.85\textwidth}{!}{%
\includegraphics{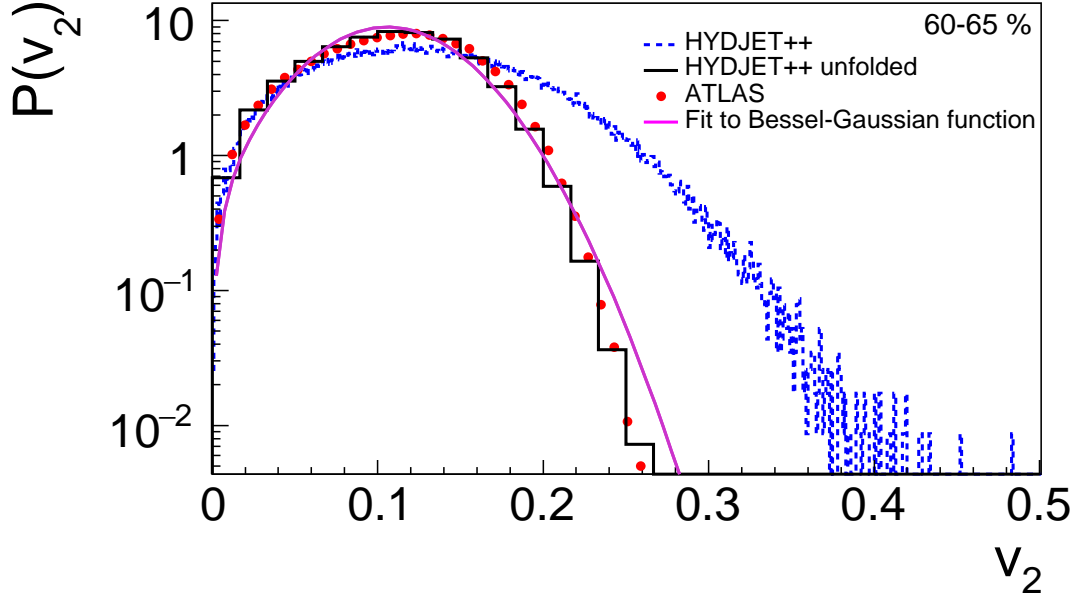}
}
\end{center}
\caption{The same as upper row of Fig.~\protect\ref{pv2} but for the 
centrality interval $60-65 \%$ only. Dashed and solid histograms present
the results for simulated HYDJET++ events before and after the unfolding 
procedure, respectively. ATLAS data are shown by full circles.
Solid curve shows the Bessel-Gauss fit to Eq.(\ref{eq:11}).
}
\label{60-65_fit}
\end{figure*}

Next question is the Gaussian-like behavior of the obtained spectra 
after the unfolding. ATLAS Collaboration reported some deviations from 
the Gaussians observed for $p(V_2)$ distributions in peripheral 
collisions \cite{Aad:2015lwa}. The last centrality bin in ATLAS analysis 
is $60-65 \%$. At this centrality the default version of HYDJET++, which
works reasonably well up to $40-45 \%$ \cite{Lokhtin:2008xi}, 
needs further fine tuning in line with other semi-phenomenological
models. Simply, the linear dependence for $v_2(b)$ becomes too
crude here. The new tuned values of the parameters are $\epsilon = 0.14$ 
and $\delta = 0.25$ (cf. with $\epsilon = 0.16$ and $\delta = 0.38$ in 
the default version). It is
worth noting that this is the only modification of the model, whereas
the ratio $\langle \epsilon \rangle / \sigma_{\epsilon}$ is kept 
constant for all
centralities in question. Figure~\ref{60-65_fit} shows the observed 
and the unfolded distributions of $p(V_2)$ obtained in HYDJET++ for 
centrality $60-65 \%$ in comparison with the ATLAS data. One can see that
the model calculations agree well with the data. The unfolded
distribution provided by HYDJET++ was also fitted to complex 
Bessel-Gauss (BG) product given by Eq.~(\ref{eq:11}). Results are plotted 
onto the simulated spectra in Fig.~\ref{60-65_fit} as well. At this 
centrality the BG fit clearly deviates from the data indicating that the
model possesses some intrinsic fluctuations which cause the distortion 
of initial Gaussians. This interesting problem definitely deserves 
further investigations.  

\begin{figure*}
\begin{center}
\resizebox{1.\textwidth}{!}{%
\includegraphics{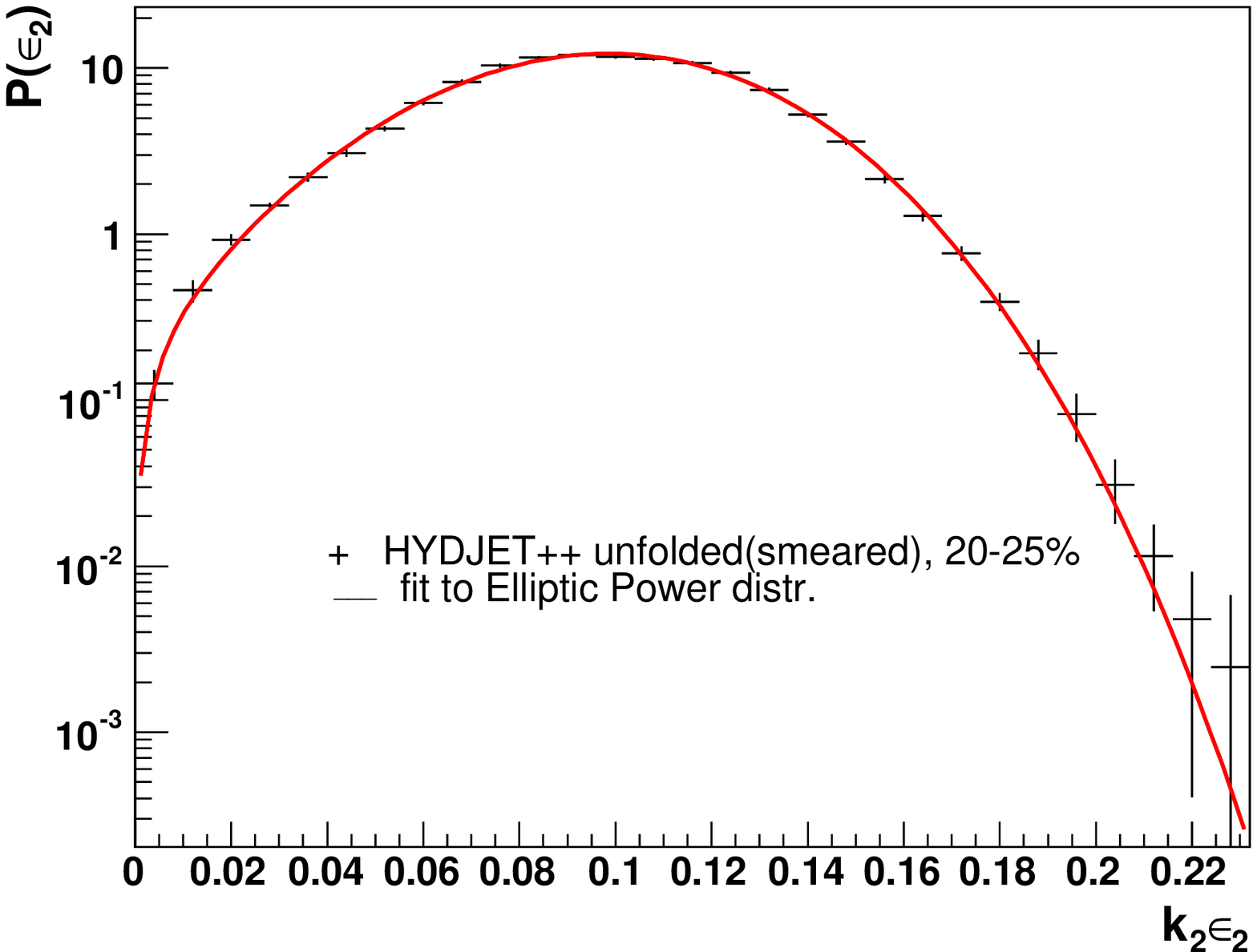}
\includegraphics{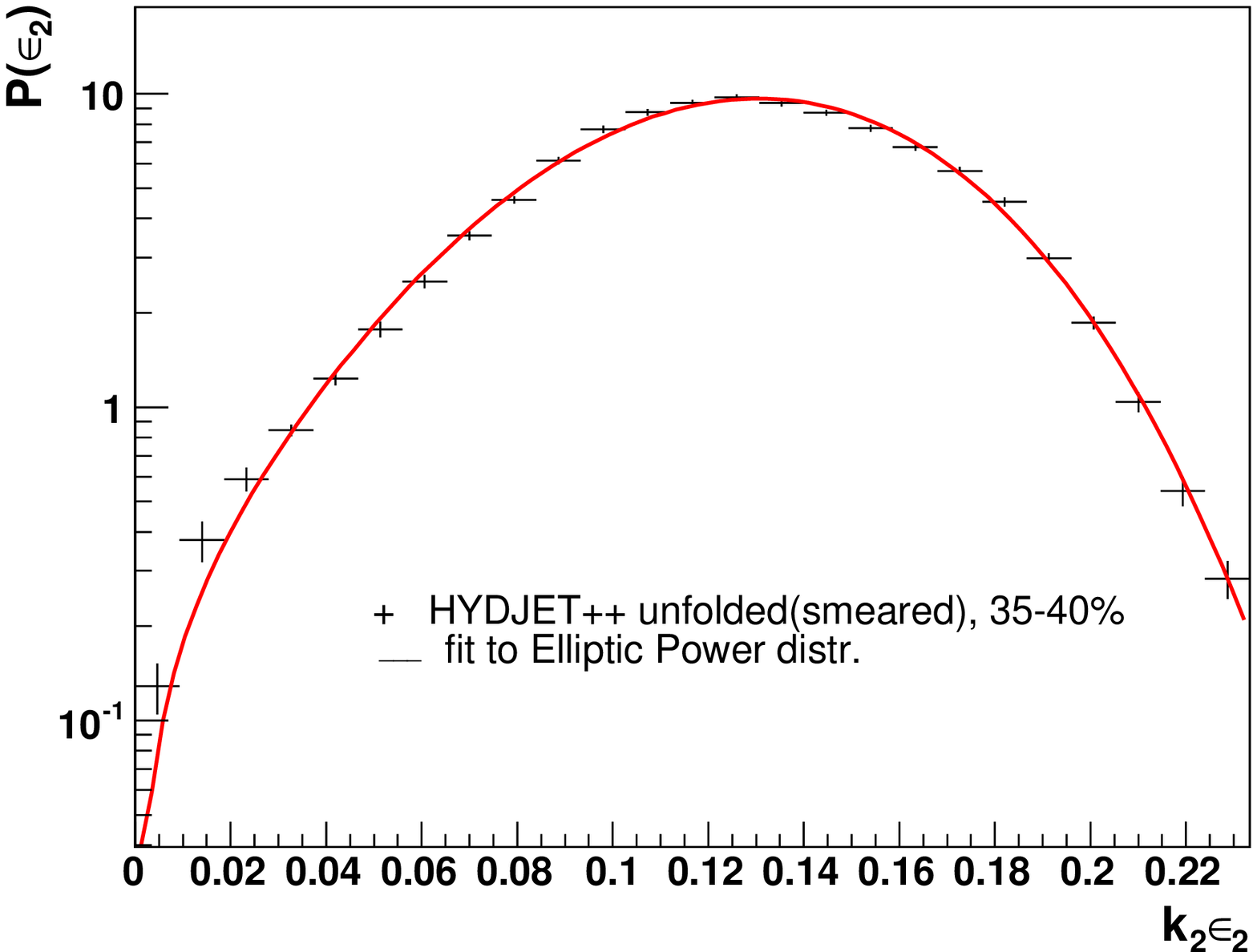}
}
\end{center}
\caption{The probability densities $P(\varepsilon_n)$ obtained in 
HYDJET++ calculations (with smearing and unfolding) of PbPb
collisions at $\sqrt{s_{\rm NN}} = 2.76$ TeV with centrality 
$20-25 \%$ and $35-40 \%$. Full lines represent the fits to the Elliptic
Power distribution~(\ref{eq:20}).}
\label{fig6}
\end{figure*}

Finally, our results can be used for the analysis of fluctuations of 
the initial anisotropy $\varepsilon_n$. This approach relies on the 
assumption of a linear response of the flow coefficient $V_n$ to the 
corresponding initial eccentricity
\begin{eqnarray}
\displaystyle
\label{eq:17}
& & V_n = k_n~\varepsilon_n \ ,
\end{eqnarray}
where $k_n$ is the response coefficient. For both elliptic and triangular 
flow this assumption works very well, as was confirmed by hydrodynamic
model calculations \cite{Qiu:2011iv,Niemi:2012aj}. The probability 
distribution of the flow is connected to the initial anisotropy
distribution via \cite{Yan:2014nsa}
\begin{eqnarray}
\displaystyle
\label{eq:18}
& & p(V_n)=\frac{d\varepsilon_n}{dV_n}~p(\varepsilon_n)~.
\end{eqnarray}
Inserting Eq.~(\ref{eq:17}) into Eq.~(\ref{eq:18}) we get
\begin{eqnarray}
\displaystyle
\label{eq:19}
& & p(\varepsilon_n) = k_n ~ p(V_n)~. 
\end{eqnarray}
Then, in \cite{Yan:2014afa} the Elliptic Power distribution was 
proposed to parametrize the eccentricity distributions
\begin{eqnarray}
\displaystyle
\label{eq:20}
\nonumber
& &
p(\varepsilon_n)=\frac{2\alpha\varepsilon_n}{\pi} (1 - 
\varepsilon_0^2)^{\alpha+1/2}
\int_0^{\pi} \frac{(1-\varepsilon_n^2)^{\alpha-1}d\varphi}   
{(1-\varepsilon_0 \varepsilon_n \cos{\varphi)}^{2\alpha+1}}~. \\
& &
\end{eqnarray}
Here the parameter $\varepsilon_0$ is approximately the mean reaction 
plane eccentricity and $\alpha$ describes the eccentricity fluctuations. 
In \cite{Yan:2014nsa} the authors fitted the ATLAS data on the elliptic
flow to Eq.~(\ref{eq:20}). It is very tempting, therefore, to fit the 
HYDJET++ generated distributions to Eq.~(\ref{eq:20}) and
compare the extracted parameters, $\alpha,\ \varepsilon_0$ and $k_2$.
The fitted curves are plotted onto the model calculations (with smearing
and unfolding procedure) of $p(\varepsilon_2)$ distributions for two 
centralities, $20 - 25\%$ and $35 - 40\%$, in Fig.~\ref{fig6}.
Extracted fit parameters are compared in Table~2 with those
obtained in \cite{Yan:2014nsa}. The agreement between the two sets is 
good, indicating that HYDJET++ quantitatively reproduces the anisotropic
flow fluctuations.

\section{Conclusions}
\label{sec_concl}

The phenomenological analysis of the event-by-event distributions of 
anisotropic flow harmonics measured in lead-lead collisions at the 
center-of-mass energy 2.76 TeV per nucleon pair has been performed 
within the two-component HYDJET++ model. The unfolding procedure was
applied to examine the simulated events, thus allowing for the direct
comparison of model calculations with the experimental data. 
To our best knowledge, this is for the first time when the
model-generated spectra are filtered by means of the unfolding
procedure and then compared directly with the LHC data obtained by the
same method. This procedure removes the non-flow effects, originating, 
e.g., from the decays of resonances and fragmentation of jets, as well 
as the finite event multiplicity effect. 

The essentially dynamical origin of the flow fluctuations in 
hydro-inspired freeze-out approach has been established. The effect is 
traced to the correlation between the momenta and coordinates of final 
particles and the velocities of hadronic fluid elements. The simple 
modification of the model via introducing the distribution over spatial 
anisotropy parameters permits HYDJET++ to reproduce both elliptic and 
triangular flow fluctuations in heavy ion collisions at the LHC energy. 
In contrast, an attempt to utilize the uniform non-Gaussian smearing 
with the same $\langle \epsilon \rangle / \sigma_\epsilon$ ratio failed 
shortly. The unfolding procedure is sensitive, therefore, to the initial 
distributions of the parameters. It should be implemented in any model 
in case of comparison with the unfolded data.

For the peripheral topologies the model calculations deviate from the
Bessel-Gaussian fit to Eq.(\ref{eq:11}) thus hinting for some intrinsic 
fluctuations in HYDJET++ which cause the distortion of initial 
Gaussians. This interesting problem deserves to be studied in the 
future.

\section*{Acknowledgments}

\begin{acknowledgement}
Discussions with G.Kh.~Eyyubova are gratefully acknowledged. We thank 
our colleagues from CMS, ALICE and ATLAS collaborations for fruitful 
cooperation. This work was supported by Russian Scientific Fund 
under grant No.14-12-00110 in a part of computer simulation of anisotropic 
flow fluctuations in PbPb collisions and the event-by-event unfolding 
analysis.

\end{acknowledgement}

%%%%%%%%%%%%%%%%%%%%%%%%%%%%%%%%%%%%%%%%%%%%%%%%%%%%

\begin{table}[hbtp]
\label{table1} 
\begin{center}
\begin{tabular}{| c | c | c | c | c | }
\hline
& Centrality & $5-10 \%$ &  $20-25 \%$ & $35-40 \%$ \\
\hline \hline
& $ \langle V_{2} \rangle $ & 0.0522 & 0.0976 & 0.1186 \\
ATLAS  & $ \sigma_{V_{2}}$ & 0.0226 & 0.0330 & 0.0422 \\ 
& $\sigma_{V_{2}}/\langle V_{2} \rangle $  & 0.433 & 0.338 & 0.356 \\ 
 \hline \hline
& $ \langle V_{2} \rangle $ & 0.0489 & 0.0853 & 0.1213 \\
HYDJET++ & $ \sigma_{V_{2}}$ & 0.0170 & 0.0231 & 0.0353 \\
& $\sigma_{V_{2}}/\langle V_{2} \rangle $  & 0.347 & 0.271 & 0.291 \\
 \hline \hline
 & $ \langle V_{2} \rangle $ & 0.0460 & 0.0823 & 0.1163 \\
HYDJET++ & $ \sigma_{V_{2}}$ & 0.0081 & 0.0095 & 0.0147 \\
unfolded & $\sigma_{V_{2}}/\langle V_{2} \rangle $  & 0.176 & 0.116 & 0.126 \\
 \hline \hline
& $ \langle V_{2} \rangle $ & 0.0580 & 0.0999 & 0.1309 \\
HYDJET++& $ \sigma_{V_{2}}$ & 0.0240 & 0.0371 & 0.0483 \\
(smeared) & $\sigma_{V_{2}}/\langle V_{2} \rangle $ & 0.414 & 0.371 & 0.369 \\
\hline \hline
HYDJET++& $ \langle V_{2} \rangle $ & 0.0552 & 0.0971 & 0.1267 \\
unfolded & $ \sigma_{V_{2}}$ & 0.0199 & 0.0320 & 0.0408 \\
(smeared) & $\sigma_{V_{2}}/\langle V_{2} \rangle $ & 0.361 & 0.330 & 0.322 \\ 
\hline \hline
& Centrality & $5-10 \%$ &  $20-25 \%$ & $35-40 \%$ \\
\hline \hline
& $ \langle V_{3} \rangle $ & 0.0281 & 0.0344 & 0.0373 \\
ATLAS  & $ \sigma_{V_{3}}$ & 0.0147 & 0.0178 & 0.0192 \\ 
& $\sigma_{V_{3}}/\langle V_{3} \rangle $  & 0.522 & 0.518 & 0.513 \\ 
 \hline \hline
& $ \langle V_{3} \rangle $ & 0.0341 & 0.0428 & 0.0555 \\
HYDJET++ & $ \sigma_{V_{3}}$ & 0.0139 & 0.0185 & 0.0268 \\
& $\sigma_{V_{3}}/\langle V_{3} \rangle $  & 0.408 & 0.432 & 0.482 \\
 \hline \hline
  & $ \langle V_{3} \rangle $ & 0.0304 & 0.0363 & 0.0425 \\
HYDJET++ & $ \sigma_{V_{3}}$ & 0.0059 & 0.0067 & 0.0142 \\
unfolded & $\sigma_{V_{3}}/\langle V_{3} \rangle $  & 0.195 & 0.184 & 0.334 \\
 \hline \hline
& $ \langle V_{3} \rangle $ & 0.0350 & 0.0438 & 0.0527 \\
HYDJET++& $ \sigma_{V_{3}}$ & 0.0178 & 0.0225 & 0.0276 \\
(smeared) & $\sigma_{V_{3}}/\langle V_{3} \rangle $ & 0.509 & 0.514 & 0.524 \\
\hline \hline 
HYDJET++& $ \langle V_{3} \rangle $ & 0.0297 & 0.0356 & 0.0387 \\
unfolded & $ \sigma_{V_{3}}$ & 0.0148 & 0.0180 & 0.0203 \\
(smeared) & $\sigma_{V_{3}}/\langle V_{3} \rangle $ & 0.499 & 0.504 & 0.526 \\
\hline
\end{tabular}
\caption{Mean values $\langle V_{n} \rangle $, widths $\sigma_{V_{n}}$ 
and ratios $\sigma_{V_{n}}/\langle V_{n} \rangle $ (n=2,3) for HYDJET++ 
simulations and ATLAS data~\cite{Aad:2013xma}.}
\end{center}
\end{table}

\begin{table}[hbtp]
\label{table2}
\begin{center}
\begin{tabular}{| c | c | c | c | }
\hline
& Centrality  &  $20-25 \%$ & $35-40 \%$ \\
\hline \hline
& $ \alpha $  & $56 \pm 6 $& $24 \pm 3$ \\
ATLAS  & $ \varepsilon_0$  & $0.25 \pm 0.02$ & $0.31 \pm 0.08$ \\
(reanalysis in \cite{Yan:2014nsa})& $k_2 $   & $0.40 \pm 0.02$ &
$0.35 \pm 0.01$ \\
 \hline \hline

& $ \alpha$  & $48 \pm 7$ & $35 \pm 3$ \\
HYDJET++& $ \varepsilon_0$  & $0.25 \pm 0.02$ & $0.30 \pm 0.01$ \\
(smeared and unfolded) & $k_2 $  & $0.40 \pm 0.02$ &  $0.36 \pm 0.02$ \\

\hline
\end{tabular}
\caption{Parameters of the fit of eccentricity distributions to
Eq.~(\protect\ref{eq:20}). See text for details.}
\end{center}
\end{table}

%%%%%%%%%%%%%%%%%%%%%%%%%%%%%%%%%%%%%%%%%%%%%%%%%%%%


\begin{thebibliography}{99}

\bibitem{Heinz:2013th} U.~Heinz, R.~Snellings, 
Ann. Rev. Nucl. Part. Sci. {\bf 63}, (2013) 123

\bibitem{Ritter:2014uca} H.G.~Ritter, R.~Stock, 
J. Phys. G {\bf 41}, (2014) 124002

\bibitem{brahms} I.~Arsene et al. (BRAHMS Collaboration), 
Nucl. Phys. A {\bf 757}, (2005) 1

\bibitem{phobos} B.B.~Back et al. (PHOBOS Collaboration), 
Nucl. Phys. A {\bf 757}, (2005) 28

\bibitem{star} J.~Adams et al. (STAR Collaboration), 
Nucl. Phys. A {\bf 757}, (2005) 102

\bibitem{phenix} K.~Adcox et al. (PHENIX Collaboration), 
Nucl. Phys. A {\bf 757}, (2005) 184

\bibitem{Aamodt:2010pa} K.~Aamodt et al. (ALICE Collaboration), 
Phys. Rev. Lett. {\bf 105}, (2010) 252302

\bibitem{ALICE:2011ab} K.~Aamodt et al. (ALICE Collaboration), 
Phys. Rev. Lett. {\bf 107}, (2011) 032301

\bibitem{Abelev2012:di} B.~Abelev et al. (ALICE Collaboration), 
Phys. Lett. B {\bf 719}, (2013) 18

\bibitem{ATLAS:2011yk} G.~Aad et al. (ATLAS Collaboration), 
Phys. Lett. B {\bf 707}, (2012) 330

\bibitem{Aad:2012bu} G.~Aad et al. (ATLAS Collaboration), 
Phys. Rev. C {\bf 86}, (2012) 014907

\bibitem{Aad:2013xma} G.~Aad et al. (ATLAS Collaboration), 
JHEP {\bf 11}, (2013) 183

\bibitem{Aad:2014eoa} G.~Aad et al. (ATLAS Collaboration), 
Eur. Phys. J. C {\bf 74}, (2014) 2982 

\bibitem{Aad:2014vba} G.~Aad et al. (ATLAS Collaboration), 
Eur. Phys. J. C {\bf 74}, (2014) 3157 

\bibitem{Aad:2015lwa} G.~Aad et al. (ATLAS Collaboration), 
Phys. Rev. C {\bf 92}, (2015) 034903

\bibitem{Chatrchyan:2012xq} S.~Chatrchyan et al. (CMS Collaboration), 
Phys. Rev. Lett. {\bf 109}, (2012) 022301

\bibitem{Chatrchyan:2012ta} S.~Chatrchyan et al. (CMS Collaboration),  
Phys. Rev. C {\bf 87}, (2013) 014902

\bibitem{Chatrchyan:2012vqa} S.~Chatrchyan et al. (CMS Collaboration), 
Phys. Rev. Lett. {\bf 110}, (2013) 042301

\bibitem{Chatrchyan:2013kba} S.~Chatrchyan et al. (CMS Collaboration), 
Phys. Rev. C {\bf 89}, (2014) 044906

\bibitem{CMS:2013bza} S.~Chatrchyan et al. (CMS Collaboration), 
JHEP {\bf 1402}, (2014) 088

\bibitem{Aad:2013sla} G.~Aad et al. (ATLAS Collaboration), 
Phys. Rev. Lett. {\bf 111}, (2013) 152301

\bibitem{ALICE:2013xna} E.~Abbas et al. (ALICE Collaboration), 
Phys. Rev. Lett. {\bf 111}, (2013) 162301

\bibitem{Abelev:2014ipa} B.~Abelev et al. (ALICE Collaboration), 
Phys. Rev. C {\bf 90}, (2014) 034904

\bibitem{Abelev:2012ola} B.~Abelev et al. (ALICE Collaboration), 
Phys. Lett. B {\bf 719}, (2013) 29

\bibitem{Chatrchyan:2013nka} S.~Chatrchyan et al. (CMS Collaboration), 
Phys. Lett. B {\bf 724}, (2013) 213

\bibitem{Aad:2014lta} G.~Aad et al. (ATLAS Collaboration), 
Phys. Rev. C {\bf 90}, (2014) 044906

\bibitem{Khachatryan:2014jra} V.~Khachatryan et al. (CMS Collaboration), 
Phys. Lett. B {\bf 742}, (2015) 200

\bibitem{Bravina:2013xla} L.V.~Bravina, B.H.~Brusheim Johansson, 
G.Kh.~Eyyubova, V.L.~Korotkikh,  I.P.~Lokhtin, L.V.~Malinina, 
S.V.~Petrushanko, A.M.~Snigirev, E.E.~Zabrodin, 
Eur. Phys. J. C {\bf 74}, (2014) 2807 

\bibitem{Lokhtin:2008xi} I.P.~Lokhtin, L.V~Malinina, S.V.~Petrushanko, 
A.M.~Snigirev, I.~Arsene, K.~Tywoniuk, 
Comput. Phys. Commun. {\bf 180}, (2009) 779

\bibitem{Eyyubova:2014dha} G.~Eyyubova, V.L.~Korotkikh, I.P.~Lokhtin, 
S.V.~Petrushanko, A.M.~Snigirev, L.V.~Bravina, E.E.~Zabrodin, 
Phys. Rev. C {\bf 91}, (2015) 064907

\bibitem{Petersen:2013vca} H.~Petersen, B.~Muller, 
Phys. Rev. C {\bf 88}, (2013) 044918

\bibitem{Qian:2013nba} W.-L.~Qian, Ph.~Mota, R.~Andrade, F.~Gardim, 
F.~Grassi, Y.~Hama, T.~Kodama,
J. Phys. G {\bf 41}, (2013) 015103

\bibitem{Huo:2013qma} P.~Huo, J.~Jia, S.~Mohapatra, 
Phys. Rev. C {\bf 90}, (2014) 024910

\bibitem{Luzum:2013yya} M.~Luzum, H.~Petersen, 
J. Phys. G {\bf 41}, (2014) 063102

\bibitem{Schenke:2013aza} B.~Schenke, P.~Tribedy, R.~Venugopalan, 
Nucl. Phys. A {\bf 926}, (2014) 102

\bibitem{Floerchinger:2014fta} S.~Floerchinger, U.-A.~Wiedemann, 
JHEP {\bf 1408}, (2014) 005

\bibitem{Yan:2014afa} L.~Yan, J.-Y.~Ollitrault, A.M.~Poskanzer, 
Phys. Rev. C {\bf 90}, (2014) 024903

\bibitem{Bhalerao:2014xra} R.S.~Bhalerao, J.-Y.~Ollitrault, S.~Pal, 
Phys. Lett. B {\bf 742}, (2015) 94

\bibitem{Song:2013gia} H.~Song, Pramana {\bf 84}, (2015) 703

\bibitem{Jia:2014pza} J.~Jia, S.~Krishnann, 
hys. Rev. C {\bf 92}, (2015) 024911

\bibitem{Niemi:2015qia} H.~Niemi, K.J.~Eskola, R. Paatelainen, 
arXiv:1505.02677 [hep-ph]

\bibitem{Rybczynski:2015wva} M.~Rybczynski, W.~Broniowski, 
arXiv:1510.08242 [nucl-th]

\bibitem{Lokhtin:2005px} I.P.~Lokhtin, A.M.~Snigirev, 
Eur. Phys. J. C {\bf 45}, (2006) 211

\bibitem{Lokhtin:2012re} I.P.~Lokhtin, A.V.~Belyaev, L.V~Malinina, 
S.V.~Petrushanko, E.P.~Rogochaya, A.M.~Snigirev, 
Eur. Phys. J. C {\bf 72}, (2012) 2045 

\bibitem{Bravina:2013ora} L.V.~Bravina, B.H.~Brusheim Johansson, 
G.Kh.~Eyyubova, V.L.~Korotkikh,  I.P.~Lokhtin, L.V.~Malinina, 
S.V.~Petrushanko, A.M.~Snigirev, E.E.~Zabrodin, 
Phys. Rev. C {\bf 89}, (2014) 024909

\bibitem{Wiedemann98} U.~Wiedemann, Phys. Rev. С {\bf 57}, (1998) 266

\bibitem{Voloshin:2007pc} S.A.~Voloshin, A.M.~Poskanzer, A.~Tang, G.~Wang, 
Phys. Lett. B {\bf 659}, (2008) 537 

\bibitem{Bhalerao:2006tp} R.S.~Bhalerao, J.-Y.~Ollitrault, 
Phys. Lett. B {\bf 641}, (2006) 260

\bibitem{Adye} T.~Adye, arXiv:1105.1160 [physics.data-an]

\bibitem{JiaMoham} J.~Jia, S.~Mohapatra, 
Phys. Rev. C {\bf 88}, (2013) 014907

\bibitem{hijing} M.~Gyulassy, X.-N.~Wang, 
Comput. Phys. Commun. {\bf 83}, (1994) 307

\bibitem{ampt} B.~Zhang, C.M.~Ko, B.A.~Li, Z.W.~Lin, 
Phys. Rev. C {\bf 61}, (2000) 067901

\bibitem{Qiu:2011iv} Z.~Qiu, U.~Heinz, 
Phys. Rev. C {\bf 84}, (2011) 024911

\bibitem{Niemi:2012aj} H.~Niemi, G.S.~Denicol, H.~Holopainen, P.~Huovinen,
Phys. Rev. C {\bf 87}, (2013) 054901 

\bibitem{Yan:2014nsa} Li~Yan, J.-Y.~Ollitrault, A.M.~Poskanzer, 
Phys. Lett. B {\bf 742}, (2015) 290

\end{thebibliography}
\end{document}